\documentclass[aps,prl,twocolumn,amsmath,amssymb,nofootinbib,showpacs]{revtex4}
\usepackage[english]{babel}
\usepackage{times}
\usepackage{latexsym}
\usepackage{graphics}
\usepackage{subfigure}
\usepackage{epsfig}
\usepackage{color}

\renewcommand{\i}{\textrm{i}}
\newcommand{\ie}{{i.e. }}
\newcommand{\eg}{{e.g. }}

\newcommand{\vect}[1]{\mathbf{#1}}

\newcommand{\Vr}{V_\textrm{\tiny R}}
\newcommand{\sigmar}{\sigma_\textrm{\tiny R}}
\newcommand{\tVr}{\widetilde{V}_\textrm{\tiny R}}

\newcommand{\kmax}{k_\textrm{max}}

\newcommand{\nBEC}{n_\textrm{c}}

\newcommand{\bv}{\hat{b}}
\newcommand{\bvc}{\hat{b}^\dagger}
\newcommand{\dens}{\hat{n}}
\newcommand{\ddens}{\delta \hat{n}}
\newcommand{\phas}{\hat{\theta}}
\newcommand{\fp}{f^+}
\newcommand{\fm}{f^-}

\newcommand{\tV}{\widetilde{V}}

\newcommand{\modVk}{\mathcal{V}_k}
\newcommand{\modCk}{\mathcal{C}_k}

\renewcommand{\section}[1]{}

\begin{document}

\title{Anderson Localization of Bogolyubov Quasiparticles in Interacting Bose-Einstein Condensates
}

\author{P.~Lugan}
\author{D.~Cl\'ement}
\author{P.~Bouyer}
\author{A.~Aspect}
\author{L.~Sanchez-Palencia}
\affiliation{
Laboratoire Charles Fabry de l'Institut d'Optique,
CNRS and Univ. Paris-Sud,
Campus Polytechnique, 
RD 128, 
F-91127 Palaiseau cedex, France}
\homepage{http://www.atomoptic.fr}

\date{\today}

\begin{abstract}
We study the Anderson localization of Bogolyubov quasiparticles in an interacting Bose-Einstein
condensate (with healing length $\xi$) 
subjected to a random potential 
(with finite correlation length $\sigmar$).
We derive analytically the Lyapunov exponent as a function of the quasiparticle 
momentum $k$ and we study the localization maximum $k_\textrm{max}$.
For 1D speckle potentials, we find that
$k_\textrm{max} \propto 1/\xi$ when $\xi \gg \sigmar$ 
while $k_\textrm{max} \propto 1/\sigmar$ when $\xi \ll \sigmar$,
and that the localization is strongest when $\xi \sim \sigmar$.
Numerical calculations support our analysis and 
our estimates indicate that the localization of the Bogolyubov quasiparticles
is accessible in 
experiments with ultracold atoms.
\end{abstract}


\maketitle

%%%%%%%%%%%%%%%%%%%%%%%%%%%%%%%%%%%%%%%%%%%%%%%%%%%%%%%%%%%%%%%%%%%%%%
%\section{Introduction}
%\label{sec:introduction}

An important issue in mesoscopic physics concerns the
effects of disorder in systems where both quantum interference 
and particle-particle interactions play crucial roles. 
Multiple 
scattering of non-interacting quantum
particles from a random potential
leads to strong Anderson localization (AL) \cite{anderson1958}, 
characterized by an exponential decay of the quantum states 
over a typical distance, the localization length. 
AL occurs for arbitrarily weak disorder in 1D and 2D, and for
strong-enough disorder in 3D \cite{gang4}.
The problem is more involved in the presence of interactions.
Strong disorder in repulsively interacting Bose gases induces 
novel insulating quantum states, such as 
the Bose \cite{boseglass} and Lifshits \cite{lugan2007} glasses.
For moderate disorder and interactions, the system forms a 
Bose-Einstein condensate (BEC) \cite{lee1990,lsp2006,lugan2007}, 
where the disorder induces
a depletion of the condensed and superfluid fractions 
\cite{fractions}
and the shift and damping of sound waves \cite{sound}.

These studies have direct applications to experiments on
liquid $^4$He in porous media \cite{helium4}, in particular
as regards the understanding of the absence of superfluidity.
Moreover, the realization of disordered gaseous BECs
\cite{lye2005,clement2005,fort2005,schulte2005,clement2006} 
has renewed the issue due to an unprecedented
control of the experimental parameters.
Using optical speckle fields \cite{goodman1975}, for instance, 
one can control the amplitude and design the correlation 
function of the random potential almost at will
\cite{clement2006}, 
opening possibilities for experimental studies of AL
\cite{kuhn,lsp2007}. Earlier studies related to localization 
in the context of ultracold atoms include dynamical 
localization in $\delta$-kicked rotors \cite{kickedrotor} and spatial diffusion 
of laser-cooled atoms in speckle potentials \cite{specklecooling}.

Transport processes in repulsively interacting BECs
can exhibit AL \cite{lsp2007,paul2007}.
However, for BECs at equilibrium,
interaction-induced delocalizing effects dominate 
disorder-induced localization, except for 
very weak interactions \cite{lee1990,lsp2006}.
The ground state of an interacting BEC at equilibrium is 
thus extended.
Beyond, one may wonder how the many-body (collective) excitations
of the BEC behave in weak disorder. 
In dilute BECs, these excitations correspond to quasiparticles 
(particle-hole pairs) described by the Bogolyubov theory \cite{bogolyubov}.
In this case, the interplay of interactions and 
disorder is subtle
and strong arguments indicate that the Bogolyubov quasiparticles (BQP) 
experience a random potential {\it screened} by the 
BEC density \cite{lee1990}.
This problem has been addressed in the idealized case of 
uncorrelated disorder (random potentials with a delta correlation function)
in Ref.~\cite{bilas2006}.

In this Letter, we present a general quantitative treatment of 
the localization of the BQPs in an interacting BEC with
healing length $\xi$ in a weak random potential with arbitrary
correlation length $\sigmar$.
For weak disorder, we introduce a transformation 
that maps rigorously the many-body Bogolyubov equations 
onto the Schr\"odinger equation for a non-interacting particle in a 
screened random potential, which we derive analytically.
We calculate the Lyapunov exponent $\Gamma_k$ (inverse localization length) as a function of the 
BQP wavenumber $k$ for 1D speckle potentials.
For a given ratio $\xi/\sigmar$, we determine the wavenumber 
$k_\textrm{max}$ for which $\Gamma_k$ is maximum: 
We find $k_\textrm{max} \sim 1/\xi$ for $\xi \gg \sigmar$, and 
$k_\textrm{max} \sim 1/\sigmar$ for $\xi \ll \sigmar$.
The absolute maximum appears for $\xi \sim \sigmar$,
so that the finite-range correlations of the disorder 
need to be taken into account.
Numerical calculations support our analysis.
Finally, possibilities to observe the AL
of BQPs in BECs placed in speckle potentials are discussed.

%%%%%%%%%%%%%%%%%%%%%%%%%%%%%%%%%%%%%%%%%%%%%%%%%%%%%%%%%%%%%%%%%%%%%
%\section{General Popov theory for a Bose gas with weak density fluctuations}
We consider a $d$-dimensional Bose gas in a potential $V(\vect{r})$ with weak repulsive short-range 
atom-atom interactions, characterized by the coupling constant $g$.
Its physics is governed by the many-body Hamiltonian
%+++++++++++++++++++++++++++++++++++++++++%
\begin{eqnarray}
\hat{H} = \int\textrm{d}\vect{r} 
\big\{ (\hbar^2/2m) [(\vect{\nabla} \phas)^2 \dens
                     + (\vect{\nabla} \sqrt{\dens})^2] \phantom{i}
\nonumber \\
        +  V(\vect{r}) \dens + (g/2) \dens^2 - \mu \dens 
\big\}
\label{eq:Hamiltonian}
\end{eqnarray}
%+++++++++++++++++++++++++++++++++++++++++%
where $m$ is the atomic mass, $\mu$ is the chemical potential, and
$\phas$ and $\dens$ are the phase and density operators, which obey
the commutation relation
$[ \dens (\vect{r}), \phas (\vect{r'}) ] = \i \delta ( \vect{r} - \vect{r'} )$. 
According to the Bogolyubov-Popov theory \cite{bogolyubov,popov1983,petrov2000}, for
small phase gradients ($\hbar^2|\nabla \theta|^2/2m \ll \mu$) and 
small density fluctuations ($\ddens \ll \nBEC$, where $\nBEC = \langle \dens \rangle$ 
and $\ddens = \dens - \nBEC$),
Hamiltonian~(\ref{eq:Hamiltonian}) can be diagonalized 
up to second order as $\hat{H} = E_0 + \sum_\nu \epsilon_\nu\ \bvc_\nu \bv_\nu$, 
where $\bv_\nu$ is the annihilation operator of the excitation (BQP) of energy 
$\epsilon_\nu$.
The many-body ground state of the Bose gas 
is a BEC with a uniform phase and 
a density governed by the Gross-Pitaevskii equation (GPE):
%+++++++++++++++++++++++++++++++++++++++++%
\begin{equation}
\mu = -\hbar^2\vect{\nabla}^2(\sqrt{\nBEC}) / 2m\sqrt{\nBEC} + V(\vect{r}) + g \nBEC (\vect{r}).
\label{eq:GPE}
\end{equation}
%+++++++++++++++++++++++++++++++++++++++++%
Expanding the density and phase in the basis of the excitations,
$\phas (\vect{r}) = [-i/2\sqrt{\nBEC(\vect{r})}] \sum_\nu [ \fp_\nu (\vect{r})\ \bv_\nu - \textrm{H.c.}]$ 
and 
$\ddens (\vect{r}) = \sqrt{\nBEC(\vect{r})} \sum_\nu [ \fm_\nu (\vect{r})\ \bv_\nu + \textrm{H.c.}]$, 
the Hamiltonian reduces to the above diagonal form 
provided that $f^\pm_\nu$ obey the Bogolyubov-de Gennes 
equations (BdGE) \cite{degennes1966}:
%+++++++++++++++++++++++++++++++++++++++++%
\begin{eqnarray}
&& \left[-(\hbar^2/2m) \vect{\nabla}^2 + V
+ \phantom{1}g\nBEC
-\mu \right] \fp_\nu 
   = \epsilon_\nu \fm_\nu
\label{eq:degennes1} \\
&& \left[-(\hbar^2/2m) \vect{\nabla}^2 + V
+ 3g\nBEC
-\mu \right] \fm_\nu
   = \epsilon_\nu \fp_\nu,
\label{eq:degennes2}
\end{eqnarray}
with the normalization 
$\int \textrm{d}\vect{r} \left[ \fp_{\nu} \fm_{\nu'}{}^{*} + \fm_{\nu} \fp_{\nu'}{}^{*} \right]
= 2\delta_{\nu,\nu '}$.
Equations~(\ref{eq:GPE})-(\ref{eq:degennes2}) form a complete set
to calculate the ground state (BEC) and excitations (BQPs) of the Bose gas, 
from which one can compute all properties 
of finite temperature or time-dependent BECs.

%%%%%%%%%%%%%%%%%%%%%%%%%%%%%%%%%%%%%%%%%%%%%%%%%%%%%%%%%%%%%%%%%%%%%
%\section{Bogolyubov modes in a 1D box with weak disorder}
Here, we analyze the properties of the BQPs
in the presence of weak disorder. 
According to Eqs.~(\ref{eq:degennes1}) and (\ref{eq:degennes2}), they are determined
by the interplay of the disorder $V$ and the BEC density background $\nBEC$.
Let $V(\vect{r})$ be a weak random potential ($\tV \ll \mu$, see below) with a vanishing average 
and a finite-range correlation function,
$C(\vect{r})=\Vr^2 c(\vect{r}/\sigmar)$, 
where $\Vr=\sqrt{\langle V^2 \rangle}$ is 
the standard deviation, and $\sigmar$ the correlation length of $V$.
As shown in 
Refs.~\cite{lugan2007,lee1990,lsp2006}, the BEC density profile 
is extended for strong-enough repulsive interactions
(\ie for $\xi \ll L$, where 
$\xi=\hbar/\sqrt{4m\mu}$ is the healing length and 
$L$ the size of the BEC).
More precisely, up to first order in $\Vr/\mu$, 
the GPE~(\ref{eq:GPE}) yields
%+++++++++++++++++++++++++++++++++++++++++%
\begin{equation}
\nBEC (\vect{r}) = [\mu - \tV (\vect{r})]/g
\label{eq:TFsmooth}
\end{equation}
%+++++++++++++++++++++++++++++++++++++++++%
where
$\tV (\vect{r}) = \int \textrm{d}\vect{r}' G_{\xi} (\vect{r}-\vect{r}') V(\vect{r}')$ 
and $G_{\xi}$, the Green function of the linearized GPE \cite{lsp2006,lugan2007}, reads 
$G_{\xi} (\vect{q})=(2\pi)^{-d/2}/[1+(|\vect{q}|\xi)^2]$ in Fourier space \cite{noteTF}. Then, 
%+++++++++++++++++++++++++++++++++++++++++%
\begin{equation}
\tV (\vect{q}) = V(\vect{q}) / [1+(|\vect{q}|\xi)^2].
\label{eq:smooth}
\end{equation}
%+++++++++++++++++++++++++++++++++++++++++%
Thus, $\xi$ is a threshold in the response 
of the density $\nBEC$ to the potential $V$, as 
$\tV (\vect{q}) \simeq V(\vect{q})$ for 
$|\vect{q}| \ll \xi^{-1}$, 
while 
$\tV (\vect{q}) \ll V(\vect{q})$ for 
$|\vect{q}| \gg \xi^{-1}$.
The potential 
$\tV (\vect{r})$ is %hence
a smoothed potential \cite{lsp2006}.
If $V$ is a homogeneous random potential, so is 
$\tV$, and 
according to Eq.~(\ref{eq:TFsmooth}), the BEC density profile $\nBEC$ is 
random but extended \cite{lsp2006,lugan2007}.

Solving the BdGEs~(\ref{eq:degennes1}) and (\ref{eq:degennes2}) 
is difficult in general because they are strongly coupled.
Yet, we show that for a weak (possibly random) potential 
$V(\vect{r})$ this hurdle can be overcome by using appropriate
linear combinations $g_\vect{k}^\pm$ of the $f_\vect{k}^\pm$ functions,
namely 
$g_\vect{k}^\pm = \pm\rho_k^{\pm 1/2}f_\vect{k}^+ +\rho_k^{\mp 1/2} f_\vect{k}^-$, 
with $\rho_k = \sqrt{1+1/(k\xi)^2}$ and $k=|\vect{k}|$.
For $V=0$, the equations for $g_\vect{k}^\pm$ are uncoupled 
[see Eqs.~(\ref{eq:decoupled1}) and (\ref{eq:decoupled2})] and
we recover the usual plane-wave solutions of wave vector 
$\vect{k}$ and energy $\epsilon_k=\rho_k (\hbar^2k^2/2m)$
\cite{bogolyubov}.
For weak but finite $V$, inserting Eq.~(\ref{eq:TFsmooth}) into 
Eqs.~(\ref{eq:degennes1}) and (\ref{eq:degennes2}), we find
%+++++++++++++++++++++++++++++++++++++++++%
\begin{eqnarray}
\frac{\hbar^2 k^2}{2m} g_\vect{k}^+
& = &
- \frac{\hbar^2}{2m} \nabla^2 g_\vect{k}^+ 
- \frac{2\rho_k \tV}{1+\rho_k^2} g_\vect{k}^- 
\nonumber \\
&& + \left[V-\frac{3+\rho_k^2}{1+\rho_k^2} \tV\right] g_\vect{k}^+
\label{eq:decoupled1} \\
- \rho_k^2 \frac{\hbar^2 k^2}{2m} g_\vect{k}^-
& = &
- \frac{\hbar^2}{2m} \nabla^2 g_\vect{k}^- 
- \frac{2\rho_k \tV}{1+\rho_k^2} g_\vect{k}^+ 
\nonumber \\
&&  + \left[V-\frac{1+3\rho_k^2}{1+\rho_k^2} \tV\right] g_\vect{k}^-.
\label{eq:decoupled2}
\end{eqnarray}
%+++++++++++++++++++++++++++++++++++++++++%
Equations~(\ref{eq:decoupled1}) and (\ref{eq:decoupled2}), which 
are coupled at most by a term of the order of $V$
[since $|\tV| \leq |V|$ and $2\rho_k/(1+\rho_k^2) \leq 1$], 
allow for perturbative approaches. 
Note that the functions $g_\vect{k}^+$ and $g_\vect{k}^-$ have
very different behaviors owing to the signs in the
left-hand-side terms in Eqs.~(\ref{eq:decoupled1}) and (\ref{eq:decoupled2}).
Equation~(\ref{eq:decoupled2}) can be solved to the lowest order in $\tVr/\mu$ 
with a Green kernel defined in Ref.~\cite{lsp2006}: 
we find 
%+++++++++++++++++++++++++++++++++++++++++%
$
g_\vect{k}^- (\vect{r}) \simeq
\frac{2/\epsilon_k}{1+\rho_k^2} \int \textrm{d}\vect{r}'
G_{\xi_k}(\vect{r}-\vect{r}') \tV (\vect{r}') g^+_\vect{k} (\vect{r}')
$,
%+++++++++++++++++++++++++++++++++++++++++%
where $\xi_k = \xi/\sqrt{1+(k\xi)^2}$.
Equation~(\ref{eq:decoupled1}) cannot be solved using the same method
because the perturbation series diverges. 
Nevertheless, from the solution for $g_\vect{k}^-$, we find that
$|g_\vect{k}^-/g_\vect{k}^+| \lesssim 
\frac{2/\epsilon_k}{1+\rho_k^2} |\tV| < |\tV|/\mu \ll 1$.
The coupling term in Eq.~(\ref{eq:decoupled1}) can thus be neglected 
to first order in $\tVr/\mu$ 
and we are left with the closed equation
%+++++++++++++++++++++++++++++++++++++++++%
\begin{equation}
-(\hbar^2/2m) \nabla^2 g_\vect{k}^+ + \modVk (\vect{r}) g_\vect{k}^+ 
\simeq (\hbar^2 k^2 / 2m) g_\vect{k}^+,
\label{eq:schrolike}
\end{equation}
where
\begin{equation}
\modVk (\vect{r}) = V(\vect{r}) - \frac{1 + 4 (k\xi)^2}{1 + 2 (k\xi)^2} \tV (\vect{r}).
\label{eq:effectiveV}
\end{equation}
%+++++++++++++++++++++++++++++++++++++++++%
Equation~(\ref{eq:schrolike}) is formally equivalent to a Schr\"odinger
equation for non-interacting bare particles with energy $\hbar^2 k^2/2m$, 
in a random potential $\modVk (\vect{r})$.
This mapping allows us to find the localization properties of the BQPs 
using standard methods for bare particles in 1D, 2D or 3D \cite{lifshits1988}.
However, since $\modVk (\vect{r})$ depends on the wave vector $\vect{k}$ itself,
the localization of the BQPs is dramatically different
from that of bare particles as discussed below.

In the remainder of the Letter, we restrict ourselves to the 1D case,
for simplicity, but also because AL is expected to be stronger in lower 
dimensions \cite{gang4}.
The Lyapunov exponent $\Gamma_k$
is a self-averaging quantity in infinite 1D systems, which can be computed
in the Born approximation using the phase formalism \cite{lifshits1988} 
(see also Ref.~\cite{lsp2007}). We get
$\Gamma_k = (\sqrt{2\pi}/8) (2m/\hbar^2 k)^2 \modCk (2k)$,
where $\modCk (q)$ is the Fourier transform of the correlation function
of $\modVk (z)$,
provided that 
$\Gamma_k \ll k$
\cite{kuhn,lsp2007,lifshits1988}.
Since $\modCk(q) \propto \langle |\modVk (q)|^2 \rangle$, the component of
$\modVk$ relevant for the calculation of $\Gamma_k$ is $\modVk (2k)$.
From Eqs.~(\ref{eq:smooth}) and (\ref{eq:effectiveV}), we find
%+++++++++++++++++++++++++++++++++++++++++%
\begin{equation}
\modVk (2k) =  \mathcal{S}(k\xi) V (2k); ~~~
\mathcal{S}(k\xi) = \frac{2(k\xi)^2}{1+2(k\xi)^2},
\label{eq:gamma0}
\end{equation}
%+++++++++++++++++++++++++++++++++++++++++%
and the Lyapunov exponent of the BQP reads
%+++++++++++++++++++++++++++++++++++++++++%
\begin{equation}
\Gamma_k =  [\mathcal{S}(k\xi)]^2 \gamma_k
\label{eq:gamma1}
\end{equation}
%+++++++++++++++++++++++++++++++++++++++++%
where
$\gamma_k = (\sqrt{2\pi}/32) (\Vr/\mu)^2 (\sigmar/k^2\xi^4) c(2k\sigmar)$
is the Lyapunov exponent for a bare particle with the same wavenumber 
$k$ \cite{lifshits1988,lsp2007}.

Let us summarize the validity conditions of the perturbative approach
presented here. It requires:
(i) the smoothing solution~(\ref{eq:TFsmooth}) to be valid
(\ie $\tVr \ll \mu$);
(ii) the coupling term proportional to $g_k^-$ in Eq.~(\ref{eq:decoupled1})
to be negligible (which is valid if $\tVr \ll \mu$); and
(iii) the phase formalism to be applicable. The latter requires
$\Gamma_k \ll k$, \ie 
$(\Vr /\mu) (\sigmar/\xi)^{1/2} \ll (k\xi)^{3/2} [1+1/2(k\xi)^2]$,
which is valid for any $k$ if $(\Vr /\mu) (\sigmar/\xi)^{1/2} \ll 1$.

Applying Eq.~(\ref{eq:gamma1}) to uncorrelated potentials
[$C(z) = 2D \delta (z)$ 
with $\sigmar \rightarrow 0$, 
$\Vr \rightarrow \infty$
and $2D=\Vr^2\sigmar \int \textrm{d}x c(x) = cst$], 
one recovers the formula for $\Gamma_k$ found in Ref.~\cite{bilas2006}.
Our approach generalizes this result to potentials 
with finite-range correlations, which proves useful 
since uncorrelated random potentials
are usually crude approximations of realistic disorder, for which
$\sigmar$ can be significantly large. 
We show below that if $\xi \lesssim \sigmar$, as \eg in the experiments 
of Refs.~\cite{lye2005,clement2005,fort2005,schulte2005,clement2006}, 
the behavior of $\Gamma_k$ versus $k$ is
dramatically affected by the finite-range correlations of the
disorder.

Let us discuss the physical content of Eqs.~(\ref{eq:schrolike}) and (\ref{eq:effectiveV}).
According to Eqs.~(\ref{eq:degennes1}) and (\ref{eq:degennes2}), the properties
of the BQPs are determined by both the bare random potential
$V$ and the BEC density $\nBEC$ in a non-trivial way.
Equation~(\ref{eq:schrolike}) makes their roles more transparent.
As the occurrence of the smoothed potential $\tV (z)$ in
Eq.~(\ref{eq:effectiveV}) is reminiscent of the presence of the meanfield 
interaction $g\nBEC$ in the
BdGEs~(\ref{eq:degennes1}) and (\ref{eq:degennes2}), it appears that the random
potential $\modVk (z)$ results from
the {\it screening} of the random potential $V (z)$ by the BEC density
background \cite{lee1990}.
More precisely, the expression~(\ref{eq:gamma0}) for the Fourier
component $\modVk (2k)$, relevant for the Lyapunov exponent of a
BQP, shows that the screening strength depends on the wavenumber $k$.
In the {\it free-particle regime} ($k \gg 1/\xi$), we find that the Lyapunov
exponent of a BQP equals that of a bare
particle with the same wavenumber ($\Gamma_k \simeq \gamma_k$), as expected. 
In the {\it phonon regime} ($k \ll 1/\xi$), the disorder is 
strongly screened and
we find $\Gamma_k \ll \gamma_k$, as in models of elastic media \cite{ishii1973}. 
Here, the localization of a BQP is strongly suppressed by the repulsive 
atom-atom interactions, as compared to a bare particle in the same bare potential.
These findings agree with and generalize the results obtained from the transfer
matrix method, which applies to potentials made of a 1D random
series of $\delta$-scatterers \cite{bilas2006}.

\begin{figure}[t!]
\begin{center}
\includegraphics[width=5.0cm]{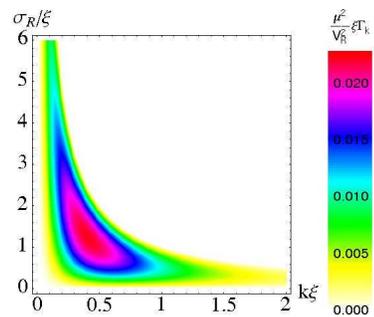}
\end{center}
\caption{(color online)
Density plot of the Lyapunov exponent of the BQPs for a 
1D speckle potential.
} 
\label{fig:gamma1}
\end{figure}

\begin{figure}[t!]
\begin{center}
\includegraphics[width=7.cm]{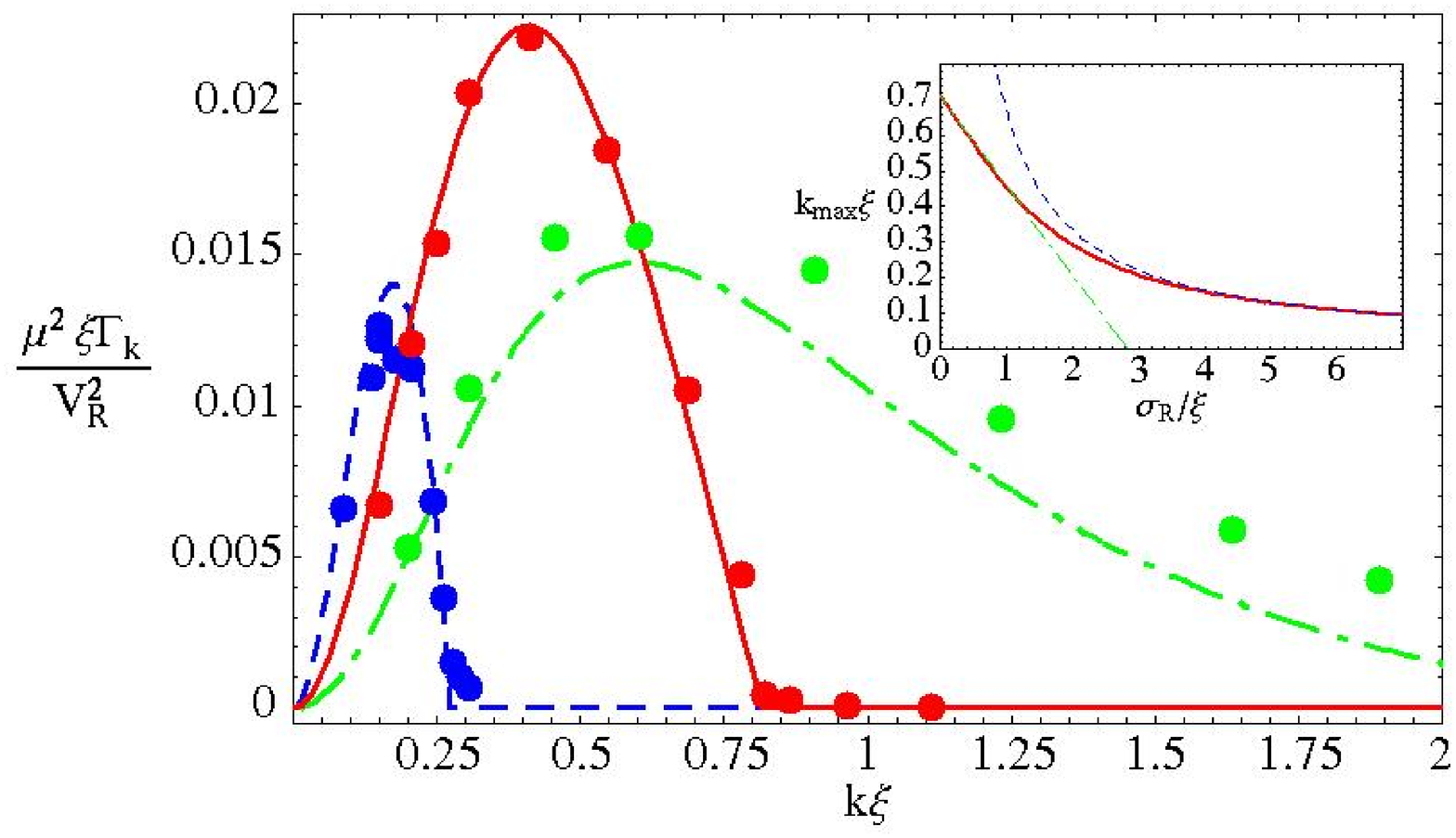}
\end{center}
\caption{(color online)
Lyapunov exponent of the BQPs for a 
1D speckle potential.
The lines correspond to Eq.~(\ref{eq:gammaspeckle})
and the points to numerical results
for $\xi=1.25\times 10^{-5}L$, $\Vr=0.075\mu$ and 
$\sigmar=\sqrt{3/2}\xi$ (solid red line),
$\sigmar=3.7\xi$ (dashed blue line),
$\sigmar=0.4\xi$ (dash-dotted green line).
Inset: localization maximum versus the ratio of 
the correlation length of the disorder to the healing length of the BEC
(the dash-dotted and dashed lines correspond to the limits
$\sigmar \ll \xi$ and $\sigmar \gg \xi$ respectively; see text).
} 
\label{fig:gamma2}
\end{figure}

%%%%%%%%%%%%%%%%%%%%%%%%%%%%%%%%%%%%%%%%%%%%%%%%%%%%%%%%%%%%%%%%%%%%%
%\section{Speckle potential}
Our approach applies to any weak random potential with a finite
correlation length. We now further examine 
the case of 1D speckle potentials used in
quantum gases \cite{lye2005,clement2005,fort2005,schulte2005,clement2006}.
Inserting the corresponding reduced correlation function,
$c(\kappa) = \sqrt{\pi/2}(1-\kappa/2)\Theta(1-\kappa/2)$
where $\Theta$ is the Heaviside function \cite{lsp2007},
into Eq.~(\ref{eq:gamma1}),
we find
%+++++++++++++++++++++++++++++++++++++++++%
\begin{equation}
\Gamma_k = \frac{\pi}{8}\left(\frac{\Vr}{\mu}\right)^2
\frac{\sigmar k^2 (1-k\sigmar)}{[1+2(k\xi)^2]^2}\Theta(1-k\sigmar)
\label{eq:gammaspeckle}
\end{equation}
%+++++++++++++++++++++++++++++++++++++++++%
which is plotted in Fig.~\ref{fig:gamma1}.
To test our general approach on the basis of this example,
we have performed numerical calculations using a direct
integration of the BdGEs~(\ref{eq:degennes1}) and (\ref{eq:degennes2}) 
in a finite but large box of size $L$.
The Lyapunov exponents are extracted from the asymptotic behavior of $\mathrm{log}[r_k(z)/r_k(z_k)]/|z-z_k|$, where $z_k$ is the localization
center and $r_k(z)$ is the envelope
of the function $g_k^+$, obtained numerically.
The numerical data, averaged over 40 realizations of the disorder, 
are in excellent agreement with formula~(\ref{eq:gammaspeckle}) as shown in Fig.~\ref{fig:gamma2}.
These results validate our approach.
It should be noted, however, that our numerical calculations return
BQP wave functions that can be strongly localized for very small
momenta $k$. This will be discussed in more details in
a future publication \cite{bogolong}.

%-----------------------------------------%
%-----------------------------------------%

%%%%%%%%%%%%%%%%%%%%%%%%%%%%%%%%%%%%%%%%%%%%%%%%%%%%%%%%%%%%%%%%%%%%%
%\section{Localization maxima}
Of special interest are the maxima of $\Gamma_k$, which denote a maximum localization of the BQPs. 
It is straightforward to show that, 
for a fixed set of parameters $(\Vr/\mu,\xi,\sigmar)$, 
$\Gamma_k$ is non-monotonic and has a single maximum, $\kmax$, in the range  $[0,1/\sigmar]$ (see Fig.~\ref{fig:gamma2}).
This contrasts with the case of bare particles, for which the 
Lyapunov exponent $\gamma_k$ decreases monotonically as a
function of $k$, provided that $c(2k\sigmar)$ decreases versus $k$ 
(which is valid for a broad class of random potentials \cite{lifshits1988}).
The existence of a localization maximum with respect to the wavenumber $k$ is thus 
specific to the BQPs and results from the strong screening of 
the disorder in the phonon regime.
In general, the value of $\kmax$, plotted in the inset of Fig.~\ref{fig:gamma2}
versus the correlation length of the disorder, 
depends on both $\xi$ and $\sigmar$.
For $\sigmar \ll \xi$, we find
$\kmax \simeq \frac{1}{\sqrt{2}\xi}\left(1-\frac{\sigmar/\xi}{2\sqrt{2}}\right)$,
so that the localization is maximum near the crossover between the phonon and the 
free-particle regimes as for uncorrelated potentials \cite{bilas2006}. 
For $\sigmar \gg \xi$ however, we find $\kmax \simeq 2/3\sigmar$, so that $\kmax$
is no longer determined by the healing length but rather
by the correlation length of the disorder, and lies deep in
the phonon regime.
For $k>1/\sigmar$, $\Gamma_k$ vanishes.
This defines an {\it effective mobility edge} due to 
long-range correlations in speckle potentials, 
as for bare particles \cite{lsp2007,notemob}.

%-----------------------------------------%
%-----------------------------------------%

Finally, let us determine the absolute localization maximum.
The Lyapunov exponent $\Gamma_k$ decreases monotonically versus 
$\xi$ and $\Vr/\mu$. However, for fixed values of $\Vr/\mu$ and $\xi$, $\Gamma_k$ has a 
maximum at $\sigmar=\sqrt{3/2}\ \xi$ and $k=\xi^{-1}/\sqrt{6}$
(see Fig.~\ref{fig:gamma1}) and we find
the corresponding localization length ($L_\textrm{max}=1/\Gamma_\textrm{max}$):
%+++++++++++++++++++++++++++++++++++++++++%
\begin{equation}
L_\textrm{max} (\xi) = (512\sqrt{6}/9\pi)(\mu/\Vr)^2 \xi.
\label{eq:gammaspecklemax}
\end{equation}
%+++++++++++++++++++++++++++++++++++++++++%
At the localization maximum, we have $\sigmar \sim \xi$, so that
the disorder cannot be modeled by an uncorrelated potential, 
and the long-range correlations must be accounted for as in our approach.
For $\sigmar=0.3\mu$m \cite{clement2006} and $\Vr=0.2\mu$, we find $L_\textrm{max} \simeq 280\mu$m,
which can be smaller than the system size in disordered, ultracold
gases \cite{clement2005,clement2006,hulet2007}.

%%%%%%%%%%%%%%%%%%%%%%%%%%%%%%%%%%%%%%%%%%%%%%%%%%%%%%%%%%%%%%%%%%%%%%
%\section{Conclusion}
In conclusion, we have presented a general treatment for the AL of BQPs 
in an interacting BEC subjected to a random potential 
with finite-range correlations.
We have calculated the Lyapunov exponents for a 1D
speckle potential and 
we have shown that the localization is strongest when $\sigmar \sim \xi$.
We have found that the localization length can be smaller than the
size of the BEC for experimentally accessible parameters.
We expect that the AL of BQPs could be observed directly, 
for instance as a broadening of the resonance lines in Bragg spectroscopy,
a well mastered technique in gaseous BECs \cite{bragg}.

%%%%%%%%%%%%%%%%%%%%%%%%%%%%%%%%%%%%%%%%%%%%%%%%%%
% \vspace{1.cm}
We thank 
M.~Lewenstein, 
G.~Shlyapnikov, 
S.~Stringari, 
and W.~Zwerger 
for stimulating discussions during the Workshop on 
Quantum Gases at the Institut Henri Poincar\'e - Centre Emile Borel.
This work was supported by the French DGA, MENRT and ANR,
and the ESF QUDEDIS program. 
The Atom Optics group at LCFIO is a member of the Institut Francilien 
de Recherche sur les Atomes Froids (IFRAF).

%%%%%%%%%%%%%%%%%%%%%%%%%%%%%%%%%%%%%%%%%%%%%%%%%%%%%%%%%%%%%%%%%%%%%%

\end{document}